# Origin of moderately volatile elements in Earth inferred from mass-dependent Ge isotope variations among chondrites

Elias Wölfer[a,*], Christoph Burkhardt[a], Francis Nimmo[b], and Thorsten Kleine[a]

[a]Max Planck Institute for Solar System Research, Justus-von-Liebig-Weg 3, 37077 Göttingen, Germany.

[b] Department of Earth and Planetary Sciences, University of California Santa Cruz, 1156 High St, Santa Cruz, CA 95064, USA

*corresponding author: woelfer@mps.mpg.de

**Abstract**

The bulk silicate Earth (BSE) is depleted in moderately volatile elements, indicating Earth formed from a mixture of volatile-rich and -poor materials. To better constrain the origin and nature of Earth's volatile-rich building blocks, we determined the mass-dependent isotope compositions of Ge in carbonaceous (CC) and enstatite chondrites. We find that, similar to other moderately volatile elements, the Ge isotope variations among the chondrites reflect mixing between volatile-rich, isotopically heavy matrix and volatile-poor, isotopically light chondrules. The Ge isotope composition of the BSE is within the chondritic range and can be accounted for as a ~2:1 mixture of CI and enstatite chondrite-derived Ge. This mixing ratio appears to be distinct from the ~1:2 ratio inferred for Zn, reflecting the different geochemical behavior of Ge (siderophile) and Zn (lithophile), and suggesting the late-stage addition of volatile-rich CC materials to Earth. On dynamical grounds it has been argued that Earth accreted CC material through a few Moon-sized embryos, in which case the Ge isotope results imply that these objects were volatile-rich, presumably because they were either undifferentiated or accreted volatile-rich objects themselves before being accreted by Earth.

# 1 Introduction

The Earth is depleted in moderately volatile elements (MVE; elements predicted to condense from a solar nebula gas between ~1250 and 650 K) compared to the primordial material of the solar accretion disk as recorded in the composition of the Sun and the most primitive meteorites, the CI (Ivuna-type) chondrites (Davis, 2006; Palme and O'Neill, 2014). This depletion is thought to reflect the accretion of Earth from a combination of volatile-poor and -rich materials (e.g., Hin et al., 2017; Mezger et al., 2021), indicating Earth incorporated chemically distinct materials from presumably distinct regions of the disk (Sossi et al., 2022). As such, identifying from where and when Earth accreted volatile-rich materials is key for understanding the origin and nature of an important subset of Earth's building blocks, and some of the underlying dynamical processes that led to changes in the provenance of Earth's accreted materials over time (Nimmo et al., 2024).

The volatile element depletion pattern of the bulk silicate Earth (BSE) strongly resembles those of carbonaceous chondrites, which has led to the suggestion that Earth accreted MVEs predominantly from carbonaceous chondrite sources (Braukmüller et al., 2019). Yet, nucleosynthetic isotope anomalies in the MVE Zn reveal that this cannot be the case. These anomalies can distinguish between non-carbonaceous (NC) and carbonaceous chondrite (CC) type meteorites, which are presumed to represent materials from the inner and outer disk, respectively (Warren, 2011; Budde et al., 2016; Kruijer et al., 2017). For Zn, the BSE has a mixed NC-CC composition (Savage et al., 2022; Steller et al., 2022; Martins et al., 2023), indicating Earth did not accrete MVEs from a single but from several sources, including materials from the inner and outer disk. However, from the Zn isotope data alone it is not possible to determine when in the accretion sequence the more volatile-rich materials were accreted.

A more detailed reconstruction of Earth's volatile accretion history is possible by examining MVEs having different geochemical characteristics. Whereas the BSE's isotopic signature of lithophile elements provides the time-averaged composition of Earth's cumulative accretion history, those of siderophile elements record only the later stages of Earth's growth because early-delivered siderophile elements were lost to the core (Dauphas, 2017). This concept has been utilized to reconstruct the accretion history of the Earth using nucleosynthetic isotope anomalies of various non-volatile lithophile and siderophile elements, showing that Earth predominantly accreted from NC materials and that a small fraction of CC materials was added during Earth's later growth stages (Budde et al., 2019; Burkhardt et al., 2021; Dauphas et al., 2024; Nimmo et al., 2024). Extending this approach to MVEs having different geochemical behavior may thus make it possible to assess the relative timing of the delivery of NC- and CC-derived MVEs to Earth. No nucleosynthetic isotope anomalies for a siderophile MVE have yet been identified, but here the case is presented that mass-dependent isotope variations in Ge indicate the late-stage delivery of volatile-rich CC bodies to Earth.

Germanium is a moderately siderophile MVE which is strongly depleted in the BSE compared to lithophile MVEs with similar condensation temperature (Palme and O'Neill, 2014). Prior studies have revealed relatively large mass-dependent Ge isotope variations between magmatic iron meteorites and ordinary chondrites (Luais, 2007, 2012; Florin et al., 2020), suggesting that different groups of meteorites may be characterized by systematically distinct Ge isotope signatures. By contrast, terrestrial igneous rocks from various geologic settings display homogeneous Ge isotope compositions, providing a good estimate for the mass-dependent Ge isotope composition of the BSE (Rouxel and Luais, 2017; Meng and Hu, 2018). Together, these characteristics make mass-dependent Ge isotope variations a promising tool for elucidating which meteorite groups, if any, contributed Ge and other MVEs to the Earth. This requires knowledge of representative Ge isotope compositions of all groups of meteorites, but until now no Ge isotope data for enstatite chondrites, whose isotopic composition for many elements resembles those of Earth's main building materials (Dauphas et al., 2014), and—with the exception of the unusual CB chondrites (Florin et al., 2021)—carbonaceous chondrites have been reported. Other MVEs such as Zn (Luck et al., 2005; Pringle et al., 2017), Te (Hellmann et al., 2020), Rb (Nie et al., 2021), K (Nie et al., 2021; Koefoed et al., 2023), Ga (Kato and Moynier, 2017), and Cd (Morton et al., 2024) exhibit systematic mass-dependent isotope variations among the different groups of carbonaceous chondrites. Provided this is also the case for Ge, such mass-dependent isotope variations may allow assessing which type of carbonaceous chondrites contributed MVEs to the Earth.

Here we present Ge isotope data for a comprehensive set of carbonaceous and enstatite chondrites. Together with data for iron meteorites and ordinary chondrites from prior studies (Luais, 2007, 2012; Florin et al., 2020), these data are used to assess the origin of MVE fractionations among the different chondrite classes and groups, and to reconstruct Earth's volatile accretion history, including the relative timing of the addition of NC- and CC-derived MVE to the growing Earth.

## 2   Samples and Methods

Twenty-three bulk chondrites (nine enstatite and 14 carbonaceous chondrites) and two USGS terrestrial basalts (BHVO-2 and BCR-2) were selected for this study (Table 1). For all chondrite samples aliquots (50-100 mg) were taken from bulk powders obtained from >2 g pieces (Hellmann et al., 2020), weighed into 15 ml Savillex PFA vials, and mixed with appropriate amounts of a $^{70}$Ge–$^{73}$Ge double spike (Wölfer et al., 2025). The methods for sample digestion, chemical separation of Ge, and mass spectrometry using a ThermoScientific Neoma MC-ICP-MS at the Max Planck Institute for Solar System Research are given in Wölfer et al. (2025). Results are reported as $\delta^{74/70}$Ge values as the permil deviation from the NIST SRM 3210a Ge standard. Repeated measurements of the NIST SRM 3120a standard provide an external reproducibility (2 s.d.) of 0.06‰ for $\delta^{74/70}$Ge. An Alfa Aesar Ge standard solution, which was measured as a secondary standard during the course of this study, is offset from the NIST SRM 3120a standard by 0.75 ± 0.07‰ (2s.d.).

The Ge isotope results for samples are reported as the mean of replicate measurements with their corresponding 95% confidence intervals. The accuracy of the Ge isotope measurements is demonstrated by repeated analyses of BHVO-2 and BCR-2, for which we determined the same $\delta^{74/70}$Ge values ($\delta^{74/70}$Ge = 0.53±0.03 for BHVO-2; $\delta^{74/70}$Ge = 0.58±0.07 for BCR-2; Table 1) as prior studies [$\delta^{74/70}$Ge = 0.53±0.12 for BHVO-2 (Escoube et al., 2012; Rouxel and Luais, 2017); $\delta^{74/70}$Ge = 0.55±0.13 for BCR-1 (Rouxel et al., 2006; Escoube et al., 2012; Luais, 2012)], which used different analytical setups (i.e. no double spike and Hydride generator instead of Aridus II).

## 3 Results

Germanium concentrations and $\delta^{74/70}$Ge values of the carbonaceous and enstatite chondrites are provided in Table 1. The carbonaceous chondrite groups exhibit variable Ge concentrations and isotopic compositions, ranging from ~10 ppm Ge and $\delta^{74/70}$Ge ≈ –1.5 for CR chondrites to ~35 ppm Ge and $\delta^{74/70}$Ge ≈ 1 for CI chondrites (Fig. 1,2). Despite these variations, samples from a given carbonaceous chondrite group exhibit similar Ge concentrations and indistinguishable $\delta^{74/70}$Ge values, indicating that parent body processes did not significantly modify the Ge isotope signatures. The $\delta^{74/70}$Ge values are correlated with Ge concentrations and matrix mass fractions inferred for each chondrite group in the order CR < CO, CV < CM < Tagish Lake, Tarda < CI (Fig. 2), demonstrating that the Ge depletion among the carbonaceous chondrites is associated with isotope fractionation towards lighter isotopic compositions (i.e., lower $\delta^{74/70}$Ge).

The EL and EH chondrites define a narrow range of $\delta^{74/70}$Ge values of between –0.4 and +0.1 and overlap with the compositions of the CV and CO chondrites (Fig. 1). These compositions partly also overlap with $\delta^{74/70}$Ge values of between –0.5 and –0.3 reported for the different groups of ordinary chondrites (Florin et al., 2020). Moreover, EH chondrites have higher Ge concentrations than EL chondrites and, albeit not resolved, tend to have slightly lighter Ge isotope compositions (Fig. 1). Similarly, Florin et al. (2020) found that the more metal-rich H chondrites are enriched in Ge over L and LL chondrites and, additionally, tend to be isotopically lighter. Importantly though, despite these small Ge isotope variations among the different groups of enstatite and ordinary chondrites, the overall $\delta^{74/70}$Ge range among these samples is much smaller than the total $\delta^{74/70}$Ge variability among the carbonaceous chondrites (Fig. 1).

## 4  Volatile element fractionations among chondrites

### *4.1  Carbonaceous chondrites*

The correlations of $\delta^{74/70}$Ge and Ge contents with matrix mass fractions among the carbonaceous chondrites strikingly resemble similar correlations for other MVEs, including Zn (Pringle et al., 2017), Te (Hellmann et al., 2020), Rb (Nie et al., 2021), and Cd (Morton et al., 2024). Consistent with this coherent behavior, there are good correlations of, for instance, Ge, Zn, and Te concentrations and isotopic compositions among the carbonaceous chondrites (Fig. 3). Of note, these elements have distinct geochemical behavior, encompassing lithophile (Rb), siderophile (Ge), and chalcophile (Zn, Cd, Te) characteristics, and cover a range of 'volatilities' with 50% condensation temperatures from ~830 K for Ge down to ~660 K for Te (Wood et al., 2019). Thus, the mechanism behind the volatile element fractionation among carbonaceous chondrites seems to be ubiquitous to all MVEs, rather than being element-specific. As pointed out in prior studies (e.g., Alexander, 2019a; Pringle et al., 2017; Hellmann et al., 2020; Nie et al., 2021; Morton et al., 2024), these systematics are best understood as the result of two-component mixing of volatile-rich, isotopically heavy CI chondrite-like matrix and a volatile-poor, isotopically light non-matrix component common to all carbonaceous chondrite groups (Fig. 3). For the MVEs, this non-matrix component is predominantly represented by chondrules. The light isotope enrichment of the chondrule component may reflect partial re-condensation during chondrule formation (Nie et al., 2021) or is a condensation signature inherited from chondrule precursors (Hellmann et al., 2020).

The composition of the non-matrix component can be inferred from the correlations of MVE concentration and isotopic composition with the mass fraction of matrix. We followed the approach of Hellmann et al. (2020), and determined the Ge concentration of the non-matrix component from the y-axis intercept of the Ge versus matrix mass fraction correlation (Fig. 2a). This approach yields a Ge concentration of 8.5±1.5 µg/g, corresponding to a CI-normalized abundance of 0.25±0.04. The $\delta^{74/70}$Ge value of the non-matrix component is then obtained from the $\delta^{74/70}$Ge versus 1/Ge correlation and is $-2.62^{+0.87}_{-1.13}$ (Fig. 2b), corresponding to an isotopic fractionation relative to the CI chondrite-like matrix of $\Delta^{74}$Ge = $3.62^{+0.87}_{-1.13}$ ‰ or ~0.9 ‰/amu. Comparison to values obtained for other MVEs reveals a much larger isotopic fractionation for Ge than for other MVEs, which typically display isotopic differences between CI chondrites and the non-matrix component of ~0.1–0.3 ‰/amu (Hellmann et al., 2020; Nie et al., 2021; Morton et al., 2024). This difference cannot be accounted for by the element's different masses, nor is it correlated with the degree of MVE depletion in the non-matrix component (Fig. 4). Thus, while all MVEs investigated thus far consistently reveal light isotopic compositions for the non-matrix component, the magnitude of the isotope fractionation appears to be element-specific and largest for Ge (Fig. 4).

Additional evidence for mixing as the main cause for isotopic variations among carbonaceous chondrites comes from the correlation of mass-dependent isotope variations of the MVEs with nucleosynthetic $^{54}$Cr isotope variations (Fig. 5b). This correlation indicates that chondrules and matrix in carbonaceous chondrites not only have distinct chemical and mass-dependent isotope compositions, but also distinct nucleosynthetic anomalies (Hellmann et al., 2023). As shown in Fig. 5b, only the CR chondrites plot off the mixing line defined by the other groups of carbonaceous chondrites. Prior studies have shown that CR chondrites contain two generations of chondrules, and it has been suggested that the offset of the CR chondrites may reflect a larger fraction of CI chondrite-like material in these chondrites which has been incorporated into the CR chondrules during repeated chondrule forming events (e.g., Marrocchi et al., 2022). However, to account for the light Ge (and Te) isotopic composition of the CR chondrites in this manner would require that during these events the CI chondrite-derived MVEs have efficiently been lost, given that CR chondrites are characterized by the lowest MVE contents and lightest isotopic composition among the carbonaceous chondrites, and plot on the $\delta^{74/70}$Ge versus 1/Ge correlation defined by the carbonaceous chondrites. When interpreted in this manner, the Ge isotope systematics of CR chondrites would support the idea that the light isotope enrichment of the non-matrix component (i.e., chondrules) for several MVEs is due to partial recondensation during chondrule formation (e.g., Nie et al., 2021).

*4.2 Enstatite chondrites*

The enstatite chondrites, like the ordinary chondrites (Florin et al., 2020), define a relatively narrow range of $\delta^{74/70}$Ge variations of ~0.5‰, which is small compared to the total ~2.5‰ $\delta^{74/70}$Ge variability among chondrites. This is in contrast to mass-dependent isotope compositions of other MVE such as Zn or Te, where enstatite and ordinary chondrites display intra-group isotope variations that are larger than the overall variations among the different chondrite classes and groups. These intra-group variations most likely reflect isotope fractionation during mobilization and redistribution of these MVEs during parent body metamorphism, where more strongly metamorphosed samples (i.e., petrological types 5 and 6) tend to show larger elemental and isotopic variations (e.g., Luck et al., 2005; Moynier et al., 2007, 2011; Hellmann et al., 2021; Braukmüller et al., 2025). For Ge, however, Florin et al. (2020) showed that ordinary chondrites of all petrologic types display rather homogeneous $\delta^{74/70}$Ge values, indicating that Ge isotope fractionation during parent body metamorphism was minor to absent. For the enstatite chondrites the effects of parent body processes are more difficult to assess, given that we analyzed only a single type 6 sample (the EL6 chondrite Khairpur). The $\delta^{74/70}$Ge value of this sample, however, is similar to those of type 3 and 4 enstatite chondrites. Thus, while a more rigorous assessment of parent body metamorphism on the Ge isotope composition of enstatite chondrites will require analyses of a more comprehensive sample set, the data of this study, together with the observation that the Ge isotope compositions of ordinary chondrites are not significantly affected by parent body metamorphism (Florin et al., 2020), suggest that the $\delta^{74/70}$Ge of type 5 and 6 enstatite chondrites will likely be similar to those of less metamorphized samples.

The disparate behavior of Ge compared to other MVEs like Zn and Te most likely results from element-specific mobilization on the chondrite parent bodies (Luck et al., 2005; Moynier et al., 2007, 2011; Hellmann et al., 2021), which in turn may be related to the different properties of an element's host phases. For instance, the major host of Ge in enstatite and ordinary chondrites is metal, while Te and Zn predominantly reside in sulfides. Given the lower thermal stability of sulfides compared to metal, it is thus conceivable that during metamorphism elements like Zn and Te were more easily mobilized from their host sulfides than Ge from its host metal. This would be consistent with the observation that enstatite and ordinary chondrites of different petrological types display larger relative elemental variations for Te and Zn than for Ge (e.g., Hellmann et al., 2021; Moynier et al., 2011; Florin et al., 2020).

The more enhanced mobility and associated isotope fractionation of elements like Zn and Te during parent body processes makes it difficult to define their bulk isotope compositions for the enstatite and ordinary chondrite parent bodies with sufficient precision that would allow assessing whether or not these compositions are distinct from those of the carbonaceous chondrites. By contrast, the smaller intra-group $\delta^{74/70}$Ge variability among enstatite and ordinary chondrites compared to the overall $\delta^{74/70}$Ge range among chondrites allows estimating the bulk $\delta^{74/70}$Ge values of their parent bodies. Based on the available data of this study, the enstatite chondrite parent bodies are characterized by $\delta^{74/70}$Ge = –0.17±0.42 (Table 1). For the ordinary chondrite parent bodies, Florin et al. (2020) reported a narrow range of $\delta^{74/70}$Ge values of between –0.51±0.09 ‰ (H chondrites) to –0.26±0.09 ‰ (LL chondrites). Thus, the enstatite and ordinary chondrite parent bodies have $\delta^{74/70}$Ge values that overlap with those of the volatile-depleted CV and CO chondrites, but are distinct from those of the volatile-rich carbonaceous chondrites (e.g., CI, CM) (Fig. 1,6). While this is consistent with the chondrule-rich nature of the enstatite and ordinary chondrites, the CR chondrites, and the non-matrix component of the carbonaceous chondrites (see above) show even lighter Ge isotope compositions. Consequently, the enstatite and ordinary chondrites plot slightly above the $\delta^{74/70}$Ge versus matrix mass fraction line defined by the carbonaceous chondrites (Fig. 5a). This offset is more clearly seen in a plot of $\delta^{74/70}$Ge versus nucleosynthetic $^{54}$Cr variations, where the carbonaceous chondrites (except CRs) define a two-component mixing line, while the enstatite and ordinary chondrites plot far off this line (Fig. 5b). Together these observations indicate that enstatite and ordinary chondrites formed from distinct precursor materials and underwent somewhat different processes of MVE fractionation than the carbonaceous chondrites.

## 5    Volatile accretion history of the Earth

The provenance of the bodies accreted by Earth is typically evaluated using nucleosynthetic isotope anomalies, which unlike mass-dependent isotope variations are not easily modified by post-accretion processes on meteorite parent bodies or the Earth itself. Nevertheless, the mass-dependent Ge isotope variations can also be used as a genetic tracer for two main reasons. First, as noted above, intra-group variations among the chondrites are smaller than the overall $\delta^{74/70}$Ge range among chondrite classes and groups (Fig. 6). Differences in $\delta^{74/70}$Ge among chondrites, therefore, predominantly reflect indigenous variations among distinct objects rather than secondary modifications on the parent bodies. Second, among terrestrial igneous rocks, volcanic rocks (tholeiitic ocean island, mid-ocean ridge, and flood basalts as well as alkali basalts), subvolcanic rocks (dolerites), plutonic rocks (chemically evolved granites, anorthosites), metamorphic rocks (serpentinites), and different mantle rocks (dunites, peridotites) all show indistinguishable $\delta^{74/70}$Ge, making it possible to define a precise $\delta^{74/70}$Ge value for the BSE of 0.60 ± 0.02 (n = 42; Rouxel et al., 2006; Escoube et al., 2012; Luais, 2012; Rouxel and Luais, 2017; Meng and Hu, 2018). Thus, the BSE and different meteorite parent bodies display well-defined and distinct $\delta^{74/70}$Ge values (Fig. 6), making it possible to assess the provenance of the bodies that delivered Ge to the BSE by comparing the $\delta^{74/70}$Ge value of the BSE to those of meteorites.

### 5.1    *Effect of core formation on $\delta^{74/70}$Ge of the bulk silicate Earth*

Although there is no significant Ge isotope fractionation during magmatic processes within the silicate Earth, the BSE's $\delta^{74/70}$Ge itself may have been modified by post-accretion isotope fractionation during chemical fractionation within the Earth. While evaporative loss from the Earth for elements as heavy as Ge can be excluded, the isotopic compositions of siderophile elements like Ge may have been modified by core formation.

Isotopic fractionation between molten metal and silicate has been identified experimentally for elements like Mo (Hin et al., 2013) and Si (Hin et al., 2014), but no such experiments have yet been performed for Ge. In ordinary chondrites and pallasites, Ge in silicates is isotopically light compared to co-existing metal, likely reflecting isotope fractionation during subsolidus cooling (Luais, 2007, 2012; Florin et al., 2020). This fractionation, however, is not a good proxy for any expected isotope fractionation between silicate and metal melts, especially because the bonding environment of Ge in solid and liquid silicate will likely be different. We will, therefore, use the experiments for Si as an analogue for Ge, given the similar valence state in silicates and chemical properties of both elements. This comparison is not ideal, given the very distinct geochemical properties of Si (predominantly lithophile) and Ge (siderophile), but given the lack of experimental data for Ge or any other siderophile element with similar properties, the comparison to Si remains the closest analogy available at this point.

Several studies have shown that silicate melts are enriched in heavy Si isotopes compared to a coexisting metallic melt, with $\Delta^{30}\text{Si}_{\text{metal-silicate}} \approx -1$ ‰ for temperatures around ~2000 K (Shahar et al., 2009, 2011; Ziegler et al., 2010; Hin et al., 2014; Poitrasson, 2017). Using the temperature dependence of the Si isotope fractionation given in Hin et al. (2014) and Shahar et al. (2011), this fractionation decreases to approximately –0.4 ‰ or –0.6 ‰, respectively, for a temperature of ~3500 K, i.e. the temperature estimated for Ge partitioning between Earth's mantle and core (assuming single stage core formation), based on the BSE's Ge abundance and experimentally determined metal-silicate partition coefficients for Ge (Righter et al., 2011). When this –0.4 ‰ (or –0.6 ‰) fractionation is normalized to the higher mass of Ge compared to Si ($\Delta^{74}\text{Ge} \approx \Delta^{30}\text{Si} \times (m_{74}-m_{70})/(m_{30}-m_{28}) \times M_{\text{Si}}/M_{\text{Ge}}$), the resulting expected fractionation for Ge is $\Delta^{74}\text{Ge}_{\text{metal-silicate}} \approx -0.3$ ‰ (or –0.5 ‰). Thus, the Ge originally delivered to BSE may have been up to ~0.3–0.5 ‰ lighter than the composition measured today. This does not prove that Ge isotopic fractionation occurred during core formation on Earth, but shows this potential effect should be taken into account when using the BSE's $\delta^{74/70}$Ge as a genetic tracer.

*5.2  Late-stage delivery of volatile-rich material to the Earth*

The BSE's pre- and post-core formation $\delta^{74/70}$Ge is within the range of meteorite values, and there are three scenarios that can reproduce the BSE's $\delta^{74/70}$Ge as mixtures of different meteorite sources: (*i*) non-carbonaceous chondrites having low $\delta^{74/70}$Ge mixed with some specific NC irons having positive $\delta^{74/70}$Ge, (*ii*) variably MVE-depleted carbonaceous chondrites (e.g., CV/CO) mixed with CI chondrites, or (*iii*) non-carbonaceous chondrites mixed with CI chondrites (Fig. 6). These scenarios differ in the fraction of CC-derived Ge in the BSE from essentially zero in the first to 100% in the second scenario. It is thus useful to assess these scenarios within the framework of current models for how the provenance of Earth's accreted materials evolved as accretion proceeded.

These models are based on nucleosynthetic isotope signatures and make use of the fact that elements with different geochemical character record different stages of a planet's growth (Dauphas, 2017). More specifically, the BSE's isotopic composition for siderophile elements records only the later stages of accretion, because earlier-accreted siderophile elements were removed to the core. This can be quantified using the parameter $x_{95}$, which is a measure of the fraction of Earth's mass after which the last 95% of an element was accreted to the mantle (Dauphas, 2017; Nimmo et al., 2024). In detail, $x_{95}$ increases with both the metal-silicate partition coefficient *D* and equilibration factor *k*, which is the fraction of an impactor's core equilibrated with Earth's mantle. Thus, higher values of $x_{95}$ are reached for more siderophile elements and a higher degree of equilibration. The $x_{95}$ values are typically calculated assuming single-stage core formation and using the difference between an element's abundance in the BSE and in chondrites (assumed to represent bulk Earth) (Dauphas, 2017; Nimmo et al., 2024).

Siderophile volatile elements like Ge are depleted in the BSE due to both core formation and volatility, and so for these elements the bulk Earth composition was estimated using the BSE's concentration of lithophile MVEs with similar condensation temperatures (see Fig. 7 for details). This approach results in an $x_{95}$ value for Ge of ~0.4.

Nimmo et al. (2024) showed that the CC fraction recorded in different elements increases with an element's $x_{95}$ value, i.e., with an element's tendency to partition into Earth's core (Fig. 7). In this model, the CC fractions are calculated as follows:

$$f_{CC} = \frac{\mu^i_{BSE} - \mu^i_{EC}}{\mu^i_{CI} - \mu^i_{EC}}$$

where $\mu^i$ refers to an isotope anomaly in either the BSE, enstatite chondrites (EC), or CI chondrites. The NC material accreted by Earth on average had an enstatite chondrite-like isotopic composition (Dauphas, 2017; Burkhardt et al., 2021; Steller et al., 2022; Dauphas et al., 2024), and so for the NC endmember it is reasonable to assume an EC-like isotopic composition. As argued in Nimmo et al. (2024), CI chondrites provide the most appropriate composition for the CC endmember, as this composition provides consistent CC fractions across a range of elements. Moreover, as shown below, CI chondrites also are a good candidate for delivering CC-derived Ge to the BSE, as they, together with some volatile-rich ungrouped chondrites like Tagish Lake and Tarda, are the only carbonaceous chondrites having a heavier $\delta^{74/70}$Ge than the BSE (Fig. 6).

Also plotted in Fig. 7 is the best fit model curve for refractory elements from Nimmo et al. (2024), showing that the variation in CC fraction with $x_{95}$ is best accounted for if Earth accreted only a small amount of CC material (about 5% by mass), which was added during the later stages of Earth's growth (the last ~10% of accretion). The inferred late-stage CC addition is predominantly based on the BSE's mixed NC-CC isotopic composition for Mo (Budde et al., 2019), but Bermingham et al. (2025) argued that the BSE's Mo is purely NC. However, we show in the Supplementary Information (Fig. S1) that the BSE's Mo most likely has a mixed NC-CC heritage, consistent with the CC fraction shown in Fig. 7.

Also shown in Fig. 7 is a model for how the CC fraction in the BSE evolved for MVEs, fitted to the CC fraction recorded for Zn and with the late-added CC-rich material enriched in Zn by a factor of 12 (Nimmo et al., 2024). For Zn, CC fractions of between ~0.3 and ~0.5 have been reported, depending on the assumed compositions of the NC and CC endmembers (Steller et al., 2022; Savage et al., 2022; Martins et al., 2023). Importantly, the mixing model used here assumes the isotopic compositions of enstatite and CI chondrites as the NC and CC endmembers, resulting in a CC fraction for Zn of 0.29±0.07 (Steller et al., 2022; Nimmo et al., 2024). With these parameters for Zn, and by assuming that the objects delivering other MVEs to Earth were enriched to the same extent as Zn, this model shows that for an $x_{95}$ value for Ge of ~0.4 from above, the CC fraction recorded by Ge should be ~0.6.

This high CC fraction is inconsistent with scenario 1 from above, in which the BSE's Ge would be purely NC. This scenario would be a viable model only if the late-stages of Earth's accretion were purely NC, but this would, as noted above, be inconsistent with the mixed NC-CC heritage of the BSE's Mo. Moreover, a purely NC origin of the BSE's Ge would imply that Earth accreted MVEs from a mix of differentiated and undifferentiated NC bodies. This is because enstatite and ordinary chondrites both have lower $\delta^{74/70}$Ge than the BSE (note that this observation holds even if the BSE's $\delta^{74/70}$Ge has been lowered by ~0.3 ‰ due to core formation and only in the extreme case of a ~0.5 ‰ fractionation would the BSE marginally overlap with the EL chondrites, see above). So, the low $\delta^{74/70}$Ge of these chondrites would have to be balanced by the addition of NC materials with more positive $\delta^{74/70}$Ge, and the only currently known such NC materials are the IC and IIAB iron meteorites. However, although these irons are not depleted in siderophile MVEs (as is evident from their near CI-chondritic Ge/Ni and Ga/Ni ratios), the silicate mantles of their parent bodies may well be depleted in lithophile

MVEs. This is because, to facilitate efficient core formation, these bodies underwent large-scale melting, such that the lithophile MVEs may have been lost by degassing from a near-surface magma ocean. Evidence that this happened comes from the eucrites and angrites, which both represent crustal samples of differentiated bodies and as is evident from for instance their low K/U ratios are strongly depleted in MVEs (Halliday and Porcelli, 2001). In such objects, the siderophile MVEs were likely not lost, because they segregated into the core prior to volatile loss. Thus, differentiated bodies may be characterized by non-chondritic ratios of lithophile to siderophile MVEs, and so would have preferentially delivered siderophile MVEs to Earth. These elements include Ga, which, while being strongly siderophile in iron meteorite parent bodies, was predominantly lithophile during core formation on Earth (Blanchard et al., 2015). However, the BSE's Ga abundance is similar to the lithophile elements Mn and Na, and only slightly elevated compared to Rb (Palme and O'Neill, 2014). Because they are lithophile, these three elements are depleted in the cores of iron meteorite parent bodies and are less and more volatile than Ga, respectively. Thus, the approximately chondritic relative abundances of Mn, Na, Ga, and Rb in the BSE suggest strongly that these elements were accreted by Earth predominantly from chondritic sources, and not from differentiated objects having sub-chondritic ratios of lithophile to siderophile MVEs. This conclusion is consistent with results of prior studies arguing that Earth accreted moderately and highly volatile elements predominantly from chondrites (e.g., Hin et al., 2017; Newcombe et al., 2023; Martins et al., 2024). Thus, unless there are NC chondrites having more positive $\delta^{74/70}$Ge values than the BSE, a purely NC origin of the BSE's Ge seems unlikely.

Taken at face value, an $x_{95}$ value of ~0.4 and CC fraction of ~0.6 are also inconsistent with scenario 2, which predicts a purely CC origin of the BSE's Ge. This scenario would be consistent with our model only if $x_{95}$ for Ge were close to one (i.e., larger than for Mo), or if the late-delivered material were more enriched in Ge than Zn. Owing to the higher condensation temperature of Ge than Zn, CI chondrites have a factor of ~2 lower Ge/Zn ratios than MVE-depleted CV and CO chondrites (Wasson and Kallemeyn, 1988). Thus, if Earth accreted CC material predominantly from MVE-depleted carbonaceous chondrites the enrichment factor would be higher for Ge than for Zn, which in turn would result in a higher predicted CC fraction at a given $x_{95}$ value. However, the $δ^{74/70}Ge$ values of the MVE-depleted CO and CV chondrites are lower than those of the BSE, and so if all the BSE's Ge were to come from carbonaceous chondrites, it would predominantly be from CI chondrites, as assumed in our model. An alternative way to accommodate a higher fraction of CC-derived Ge in the BSE is to assume that the actual $x_{95}$ for Ge is higher than calculated above. Since $D$ is lower for Ge than for Mo, a higher $x_{95}$ value could only be reached by assuming a higher value of $k$ for Ge than for Mo. This in turn would only be possible if the objects that delivered the MVEs including Ge underwent more efficient mantle-core re-equilibration (higher $k$) than the objects that were essentially MVE-free. Although somewhat ad hoc, such a scenario is difficult to exclude entirely, because the objects delivering the MVEs likely were undifferentiated bodies (see above), for which the degree of re-equilibration was probably higher than for differentiated objects.

In the third scenario, the BSE's Ge derives from a mixture of NC and volatile-rich CC chondrites, which is consistent with the mixed NC-CC heritage predicted by our model (Fig. 7). To assess this scenario more quantitatively, we calculate the fraction of CC-derived Ge in the BSE by mass balance using equation (1) from above, but using $δ^{74/70}Ge$ values instead of isotope anomalies. As noted above, the NC material accreted by Earth on average had an enstatite chondrite-like isotopic composition, and so it is reasonable to assume the $δ^{74/70}Ge$ of enstatite chondrites as the composition of the NC endmember delivering Ge to the BSE. Since the enstatite chondrites have lower $δ^{74/70}Ge$ values than the BSE, the CC endmember must be characterized by $δ^{74/70}Ge$ values more positive than the BSE. This leaves volatile-rich carbonaceous chondrites such as Tagish Lake and Tarda, and the CI chondrites as the most likely candidates for delivering CC-derived Ge to the BSE. Using the $\delta^{74/70}$Ge values of enstatite chondrites and CI chondrites, and $\delta^{74/70}$Ge = 0.60±0.02 for the BSE, equation (1) returns a CC fraction for Ge of 0.64±0.16 (2σ), in remarkable agreement with the predictions of our model, which is also based on CC fractions calculated using enstatite and CI chondrites as the isotopic endmembers (Fig. 7). Nevertheless, as noted above, the pre-core formation $\delta^{74/70}$Ge of the BSE may have been lower than the measured present-day value, resulting in a lower calculated CC fraction. However, in this case it would also be possible that the CC-derived Ge does not exclusively derive from CI chondrites but also from other chondrites having lower $\delta^{74/70}$Ge, in which case the calculated CC fraction would increase again.

In summary, despite uncertainties in the effect of core formation on the BSE's $\delta^{74/70}$Ge and the exact $\delta^{74/70}$Ge values of the carbonaceous chondrites accreted by Earth, the Ge isotope systematics in most cases indicate a relatively high fraction of CC-derived Ge in the BSE, consistent with the mixed NC-CC heritage of Earth's MVEs based on nucleosynthetic Zn isotope anomalies. In our preferred model, which is based on the measured $\delta^{74/70}$Ge of the BSE and assumes that the NC and CC materials accreted by Earth had the Ge isotope composition of enstatite and CI chondrites, respectively, the CC fractions for Ge (0.64±0.16) is larger than for Zn (0.29±0.07). This difference can be readily understood as reflecting the late-stage accretion of volatile-rich CC bodies to Earth, which left a stronger imprint on the isotopic composition of siderophile Ge compared to lithophile Zn. This interpretation is consistent with the late-stage addition of CC material inferred from nucleosynthetic Mo isotope signatures.

Nimmo et al. (2024) have shown that the late-stage addition of CC bodies to Earth, combined with the lower CC fraction in Mars compared to Earth (Kleine et al., 2023; Paquet et al., 2023), is best accounted for in a model where Earth accreted CC material predominantly from a few Moon-sized embryos. The Ge isotope data of this study and previous Zn measurements both indicate that these bodies were volatile-rich (Fig. 7), which in turn implies that these bodies were either undifferentiated, did not lose MVEs because they were too large, or accreted MVEs through undifferentiated objects themselves before being accreted to Earth.

*5.3 Implications for the late veneer*

Our finding of the late-stage addition of volatile-rich CC materials to the Earth raises the question of whether such materials also contributed to the late veneer, the final 0.5% of Earth's accretion following the cessation of core formation. Among the MVEs in the BSE, Te is thought to entirely derive from the late veneer because it is even more strongly depleted than the highly siderophile elements, all of which were contributed by the late veneer (e.g., Wang and Becker, 2013). Based on elemental ratios of S, Se, and Te, Wang and Becker (2013) argued that the BSE is distinct from enstatite chondrites but similar to some volatile-rich carbonaceous chondrites, especially the CM chondrites, suggesting a CC-dominated late veneer. However, others have argued that the S-Se-Te elemental systematics of both the BSE and chondrites may have been modified by processes on the meteorite parent bodies and in Earth's mantle and, hence, are insufficiently known to constrain the origin of the late veneer (Labidi et al., 2013, 2018; Hellmann et al., 2021). Also, the Re-Os isotope systematics of Earth's mantle (Meisel et al., 1996) and nucleosynthetic Ru isotope anomalies in lunar impactites (Worsham and Kleine, 2021) suggest a predominantly NC late veneer. Moreover, the results of this study, together with prior results on nucleosynthetic Zn isotope anomalies, show that the MVEs in the BSE are unlikely to derive from a single source, but more likely are contributed from a mix of NC and CC bodies. This raises the question of whether mass-dependent Te isotope variations, similar to Ge, can be used to estimate the fraction of CC-derived Te in the BSE.

As shown in Fig. 3b, carbonaceous chondrites exhibit correlated $\delta^{74/70}$Ge and $\delta^{128/126}$Te variations. Enstatite chondrites also plot on this correlation, but we note that for Te, enstatite chondrites display large isotope variations and so their bulk $\delta^{128/126}$Te is not as well-defined as for Ge. Nevertheless, Hellmann et al. (2021) have shown that enstatite chondrites of petrologic types 3 and 4 show quite homogeneous Te isotope compositions, and here we assume that this composition represents the bulk $\delta^{128/126}$Te of enstatite chondrite parent bodies. As shown in Fig. 8, the BSE plots on the $\delta^{74/70}$Ge–$\delta^{128/126}$Te correlation defined by the carbonaceous and enstatite chondrites, which, taken at face value, suggests similar CC fractions for Ge and Te in the BSE. Consistent with this, the fraction of CC-derived Te in the BSE calculated using equation (1) is 0.57±0.40 (2σ), which is essentially identical to the fraction of CC-derived Ge (0.64±0.14, see above). However, the uncertainty on the CC fraction for Te is quite large, reflecting the small overall $\delta^{128/126}$Te range among the chondrites and the fact that the BSE's $\delta^{128/126}$Te partly overlaps with the values of both CI and enstatite chondrites. Thus, even for different fractions of CC-derived Ge and Te, the BSE would plot close to the $\delta^{74/70}$Ge–$\delta^{128/126}$Te correlation defined by chondrites. This makes any estimate of the contribution of CC material, if any, to the late veneer based on the Te isotope systematics quite uncertain.

## 6    Conclusions

This study reports the first mass-dependent Ge isotope data for carbonaceous and enstatite chondrites, which, when combined with data for ordinary chondrites from a prior study (Florin et al., 2020), allows for a comprehensive assessment of moderately volatile element fractionations among chondritic meteorites, and for identifying the nature and origin of the objects delivering these elements to Earth. The Ge isotope variations among the carbonaceous chondrites reflect mixing between volatile-rich, isotopically heavy and CI chondrite-like matrix with volatile-poor and isotopically light chondrules or chondrule precursors. These systematics are similar to those previously observed for mass-dependent isotope variations of other moderately volatile elements, such as Zn (Pringle et al., 2017) or Te (Hellmann et al., 2020).

Consistent with their chondrule-rich nature, enstatite chondrites, like ordinary chondrites (Florin et al., 2020), have lighter Ge isotope compositions than the more volatile-rich carbonaceous chondrites. However, differences in nucleosynthetic isotope signatures among these chondrites indicate they formed from different precursor materials.

In our preferred model, the Ge isotope composition of the BSE reflects a mixture of materials having the isotopic composition of enstatite and CI chondrites, respectively, and records a larger CC fraction (~2:1) than inferred from nucleosynthetic isotope anomalies in Zn (~1:2). This difference suggests the late-stage accretion of volatile-rich CC bodies to Earth, which left a stronger imprint on the isotopic composition of siderophile Ge compared to the lithophile Zn, and is consistent with the BSE's mixed NC-CC Mo isotopic composition. Prior studies have argued that the late-stage addition of CC bodies implies that these bodies were approximately Moon-sized or larger, in order to avoid making Mars too CC-rich and because smaller, planetesimal-sized objects are expected to have been accreted earlier (Kleine et al., 2023; Nimmo et al., 2024). Within this framework, the Ge isotope data of this study indicate that these Moon-sized embryos were volatile-rich, suggesting they may have been either undifferentiated, did not lose MVEs because they were too large, or accreted MVEs themselves prior to being accreted by Earth.

**Appendix A. Supplementary Material**

Supplementary material related to this article is attached below.

**Data Availability**

All data are available in the manuscript and Supplementary Material.

**Acknowledgments**

We thank the NASA, the Field Museum, and the IPGP for providing meteorite samples. We also thank two anonymous reviewers for their constructive comments, Olivier Mousis for efficient editorial handling, and Fred Moynier for helpful discussion. This work was funded by the German Research Foundation – Project-ID 263649064 – TRR 170 and the European Research Council Advanced Grant HolyEarth (grant no. 101019380). FN acknowledges support from NSF-CSEDI-2054876.

**Table 1:** Ge isotopic and concentration data for chondrites and terrestrial samples.

| Sample | Classification | Ge [a] (μg/g) (±2σ) | N | $\delta^{74/70}$Ge (± 95% CI) |
|---|---|---|---|---|
| *Carbonaceous chondrites* | | | | |
| Orgueil | CI1 | 34.6 ± 0.4 | 7 | 1.00 ± 0.04 |
| Tarda | C2-ung. | 24.1 ± 0.3 | 6 | 0.79 ± 0.04 |
| Tagish Lake | C2-ung. | 27.5 ± 0.3 | 7 | 0.71 ± 0.04 |
| | | | | |
| MET 01070,44 | CM1 | 26.0 ± 0.3 | 7 | 0.48 ± 0.02 |
| Murchison | CM2 | 23.8 ± 0.3 | 8 | 0.51 ± 0.05 |
| Jbilet Winselwan | CM2 | 25.1 ± 0.3 | 5 | 0.49 ± 0.05 |
| **Mean CM chondrites** | | **24.9 ± 2.8** | **3** | **0.49 ± 0.02** |
| | | | | |
| DaG 136 | CO3 | 18.3 ± 0.2 | 8 | –0.28 ± 0.04 |
| NWA 5933 | CO3 | 18.9 ± 0.2 | 5 | –0.42 ± 0.07 |
| NWA 6015 | CO3 | 19.0 ± 0.2 | 8 | –0.17 ± 0.05 |
| Kainsaz | CO3 | 19.1 ± 0.2 | 5 | –0.26 ± 0.04 |
| **Mean CO chondrites** | | **18.8 ± 0.6** | **4** | **–0.27 ± 0.13** |
| | | | | |
| Vigarano | CV3 | 15.4 ± 0.2 | 5 | –0.07 ± 0.05 |
| Allende (MS-A) | CV3 | 16.8 ± 0.2 | 5 | –0.02 ± 0.05 |
| **Mean CV chondrites** | | **16.1 ± 2.1** | **3** | **–0.05 ± 0.04** |
| | | | | |
| Acfer 139 | CR2 | 9.38 ± 0.10 | 8 | –1.46 ± 0.05 |
| NWA 801 | CR2 | 10.1 ± 0.1 | 5 | –1.47 ± 0.05 |
| **Mean CR chondrites** | | **9.74 ± 1.01** | **2** | **–1.47 ± 0.04** |
| | | | | |
| *Non-carbonaceous chondrites* | | | | |
| Sahara 97072 | EH3 | 40.9 ± 0.5 | 5 | –0.42 ± 0.05 |
| LAR 12001,15 | EH3 | 39.8 ± 0.2 | 10 | –0.36 ± 0.03 |
| GRO 95517,42 | EH3 | 42.1 ± 0.2 | 10 | –0.36 ± 0.03 |
| MIL 07028,24[b] | EH3 | 45.9 ± 0.3 | 10 | 0.04 ± 0.04 |
| LAR 06252 | EH3 | 39.6 ± 0.2 | 10 | –0.31 ± 0.05 |
| **Mean EH3 chondrites**[c] | | **41.6 ± 5.1** | **4** | **–0.36 ± 0.05** |
| | | | | |
| PCA 91020,59 | EL3 | 31.7 ± 0.3 | 5 | 0.11 ± 0.04 |
| MAC 02837,45 | EL3 | 22.3 ± 0.1 | 11 | 0.01 ± 0.05 |
| MAC 02747 | EL4 | 21.8 ± 0.1 | 11 | 0.00 ± 0.03 |
| Khairpur | EL6 | 21.6 ± 0.2 | 5 | –0.25 ± 0.05 |
| **Mean EL chondrites** | | **24.4 ± 9.8** | **4** | **–0.01 ± 0.21** |
| | | | | |
| **Mean Enstatite chondrites (2 s.d.)** | | **34.0 ± 19.5** | **9** | **–0.17 ± 0.42** |
| | | | | |
| *Terrestrial standards* | | | | |
| BHVO-2[c] | Basalt | 1.63 ± 0.02 | 6 | 0.53 ± 0.03 |
| BCR-2[c] | Basalt | 1.57 ± 0.02 | 4 | 0.58 ± 0.07 |
| **Mean Bulk Silicate Earth** | | **1.60 ± 0.09** | **2** | **0.54 ± 0.03** |

Ge concentration and $\delta^{74/70}$Ge for individual samples are reported as the mean of pooled measurements. Group means represent weighted means of individual samples of each group, calculated using IsoplotR. N: number of Ge isotope analyses.

[a] Ge concentrations in this study have been determined by isotope dilution.
[b] $\delta^{74/70}$Ge of MIL 07028,24 not included in calculation of weighted mean $\delta^{74/70}$Ge of EH3 chondrites.
[c] Processed and measured together with the chondrites of this study, but published already in Wölfer et al. (2025).

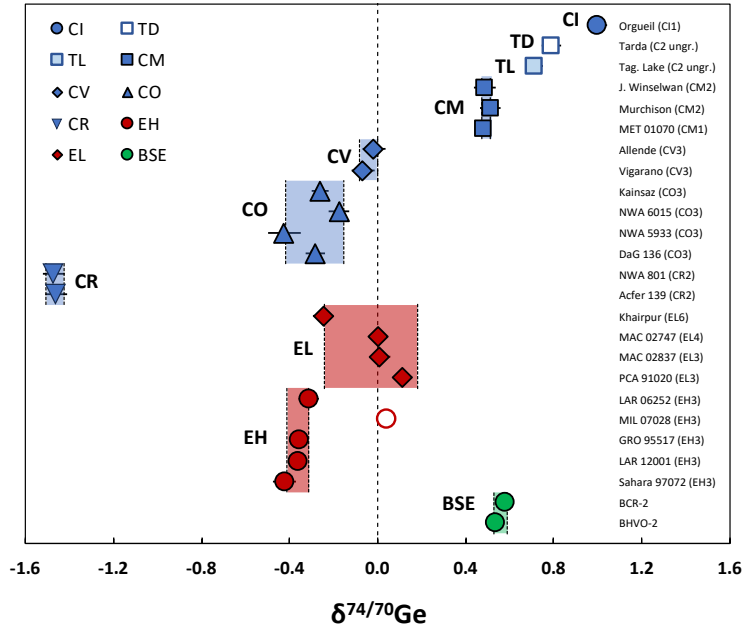

**Fig. 1:** $\delta^{74/70}$Ge variations among carbonaceous (blue) and enstatite chondrites (red) of this study. Results for two terrestrial basalts (green) analyzed in this study are shown for comparison and overlap with the composition of the BSE. Boxes indicate the weighted mean $\delta^{74/70}$Ge of each group, calculated using IsoplotR. The $\delta^{74/70}$Ge of MIL 07028,24 (open red circle) was excluded from the weighted mean of the EH chondrites.

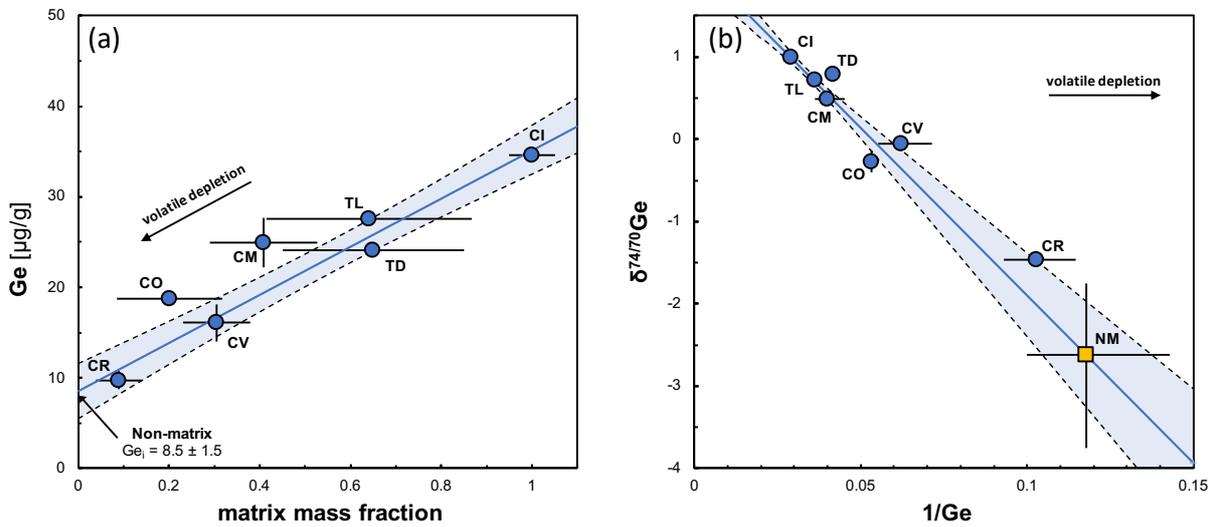

**Fig. 2:** Diagrams of (a) Ge concentration vs. matrix mass fraction and (b) $\delta^{74/70}$Ge vs. 1/Ge for carbonaceous chondrites (blue circles) and the inferred non-matrix component (orange square). Matrix mass fractions are from Hellmann et al. (2020) and $\epsilon^{54}$Cr values are from the compilations of Hellmann et al. (2020, 2023) and Spitzer et al. (2020). Linear regressions and error envelopes (95% CI) were calculated using Isoplot (Ludwig, 2012). Tarda was excluded from the regression in (b). TD: Tarda. TL: Tagish Lake.

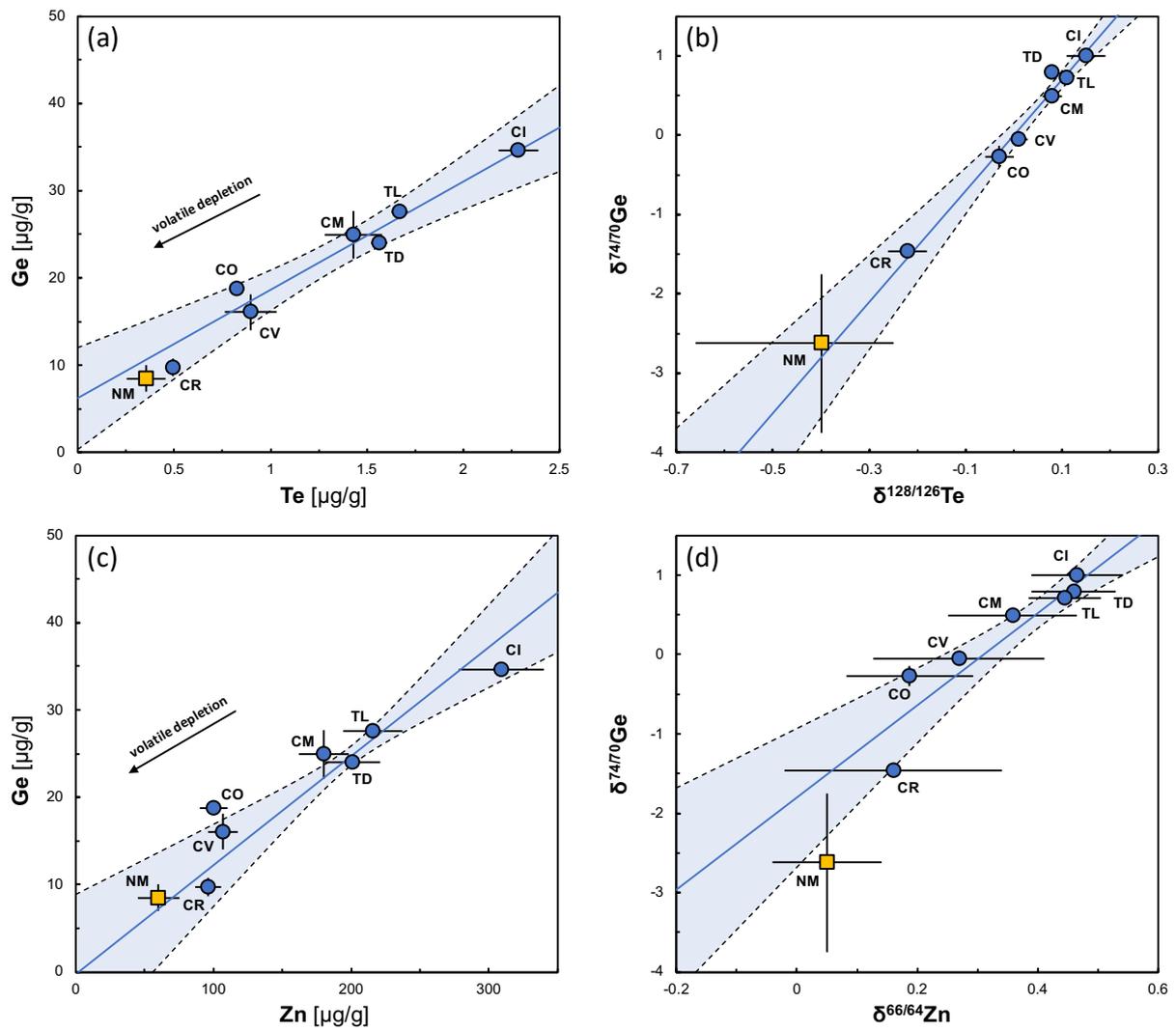

**Fig. 3:** Diagrams of (a) Ge vs. Te, (b) $\delta^{74/70}$Ge vs. $\delta^{128/126}$Te, (c) Ge vs. Zn, and (d) $\delta^{74/70}$Ge vs. $\delta^{66/64}$Zn for carbonaceous chondrites (blue circles) and the inferred non-matrix component (orange square). Tellurium data are from Hellmann et al. (2020), and Zn data are from Luck et al. (2005), Barrat et al. (2012), Pringle et al. (2017), Alexander (2019a, b), Mahan et al. (2018), Paquet et al. (2022), Morton et al. (2024), and Nie et al. (2021). As in Fig. 2, linear regressions were calculated using Isoplot (Ludwig, 2012). TD: Tarda. TL: Tagish Lake.

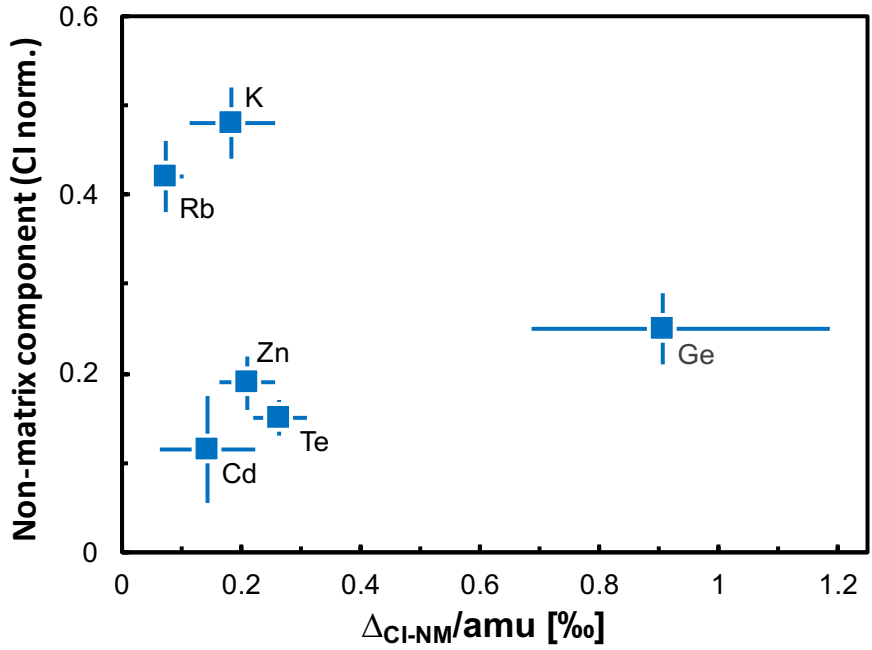

**Fig. 4**: Mass-dependent isotope fractionation between CI chondrite-like matrix and non-matrix component ($\Delta_{CI-NM}$/amu [‰]) versus CI-normalized abundances in non-matrix component. Data for Ge are from this study; data for K, Rb, Zn, and Te are from Nie et al. (2021), and data for Cd are from Morton et al. (2024). Note that the isotopic fractionation for Ge is much larger than for any of the other moderately volatile elements.

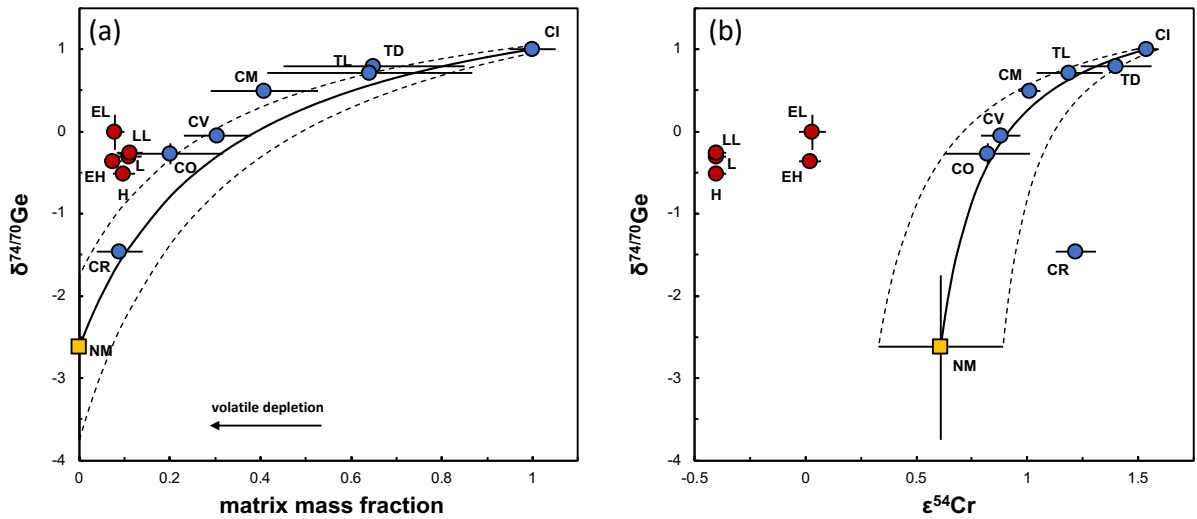

**Fig. 5:** Diagrams of (a) $\delta^{74/70}$Ge vs. matrix mass fraction and (b) $\delta^{74/70}$Ge vs. $\varepsilon^{54}$Cr for carbonaceous (blue circles) and non-carbonaceous (red circles) chondrites. The non-matrix component (orange square) is also shown. Germanium data for ordinary chondrites (H, L, LL) from Alexander (2019) and Florin et al. (2020), matrix mass fractions are from Alexander (2019a, b) and Hellmann et al. (2020), and $\varepsilon^{54}$Cr values are from the compilations of Hellmann et al. (2020, 2023) and Spitzer et al. (2020). Black lines and associated error envelopes are mixing lines between CI chondrite-like matrix and the non-matrix component. TD: Tarda. TL: Tagish Lake.

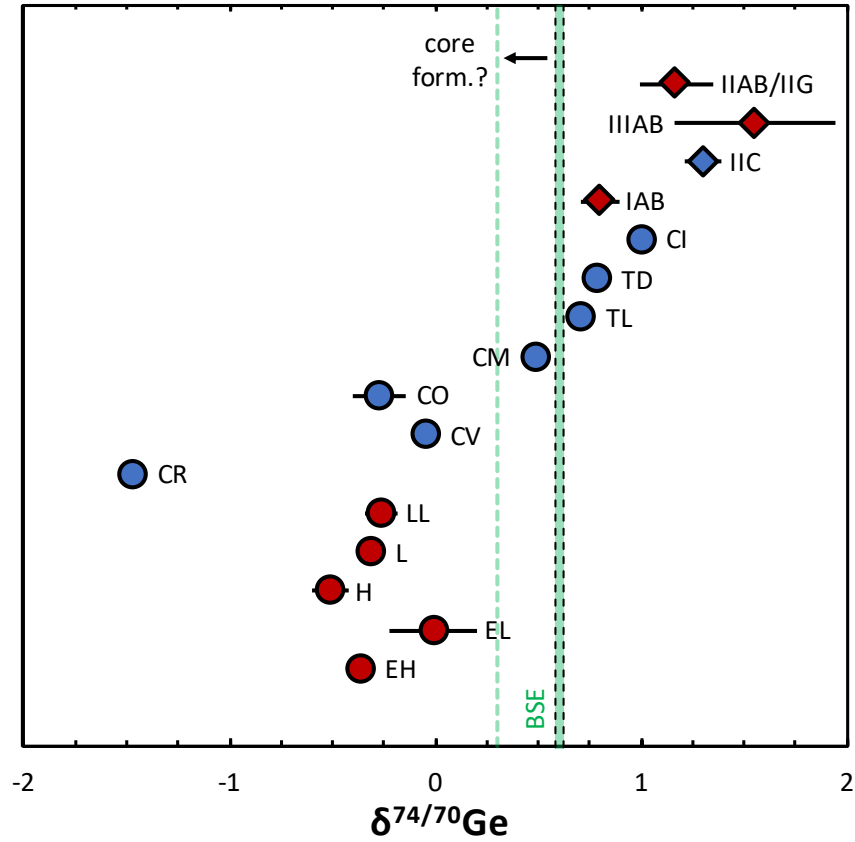

**Fig. 6:** $\delta^{74/70}$Ge variations among carbonaceous (blue) and non-carbonaceous (red) meteorites. Chondrites are shown as circles, iron meteorites as diamonds. Green bar indicates the $\delta^{74/70}$Ge value of the bulk silicate Earth (BSE) with the dotted green line showing the potential effect of core formation on the BSE's $\delta^{74/70}$Ge (see section 5.1 for details). Composite points of each meteorite group represent the weighted mean of individual samples (Table S2). TD: Tarda. TL: Tagish Lake.

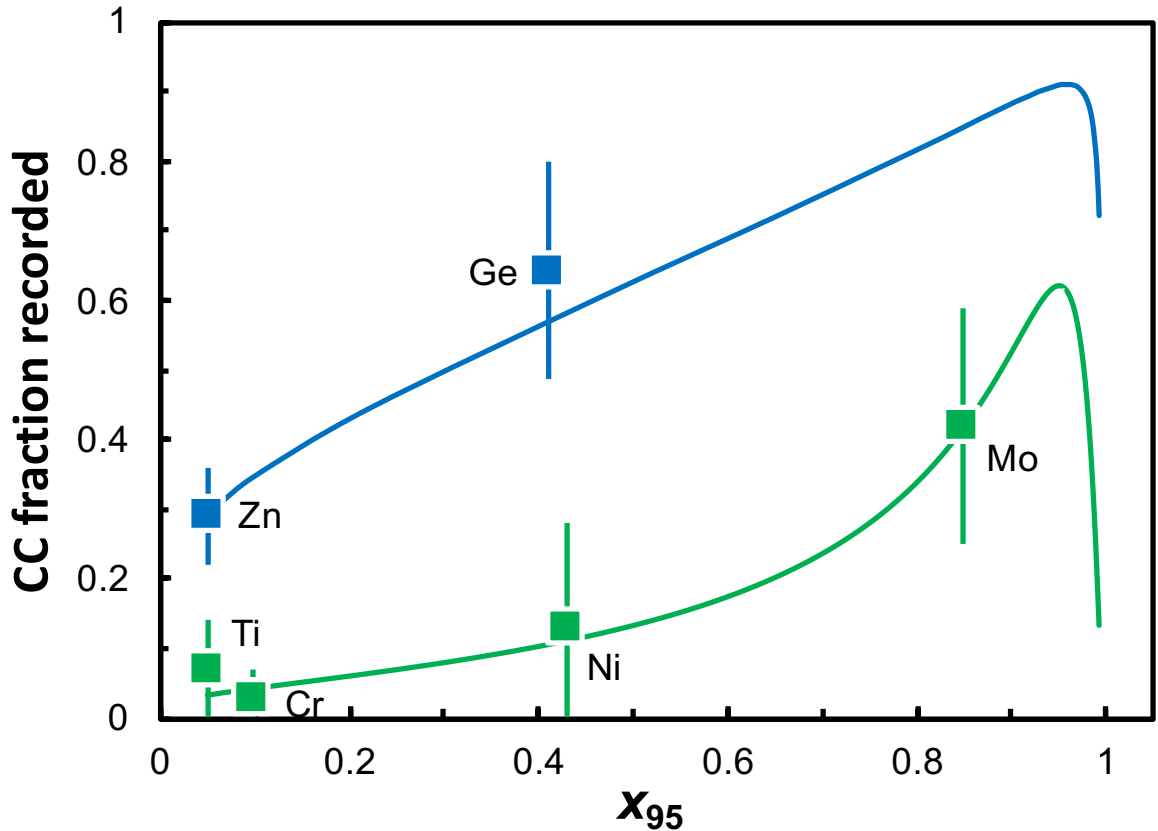

**Fig. 7:** CC fractions recorded in different elements versus $x_{95}$ (for definition see section 5.2). Green symbols represent non-volatile elements, blue symbols the MVEs Zn and Ge. Green solid line shows the best fit model from Nimmo et al. (2024) to reproduce the variable CC fractions for non-volatile elements. Blue line is the same model fitted to account for the CC fraction recorded in the MVE Zn [see Fig. 5 in Nimmo et al. (2024)] where the late-delivered CC-rich material is enriched in Zn by a factor of 12 relative to the early-delivered material. Also shown is the CC fraction for Ge determined in this study (for scenario 3, see section 5.2) and the $x_{95}$ value for Ge calculated as follows: For the bulk Earth abundance we take the lithophile MVEs Na and Rb as proxies, which have slightly higher and lower condensation temperatures than Ge. This results in a Ge abundance of ~10 ppm for the bulk Earth. For the BSE we take a Ge concentration of 1.2 ppm (Palme and O'Neill, 2014). Assuming single stage core formation, a $D$ value of 24 is calculated. Assuming $k = 0.2$ (the value assumed for the other elements shown in this plot; Nimmo et al., 2024), we calculate an $x_{95}$ value of ~0.4. For these set of parameters, the BSE's Ge isotope signature is consistent with the late-stage delivery of CC material to the Earth. Note that the blue line is based on Zn alone, but predicts the inferred CC fraction of Ge in the BSE quite well.

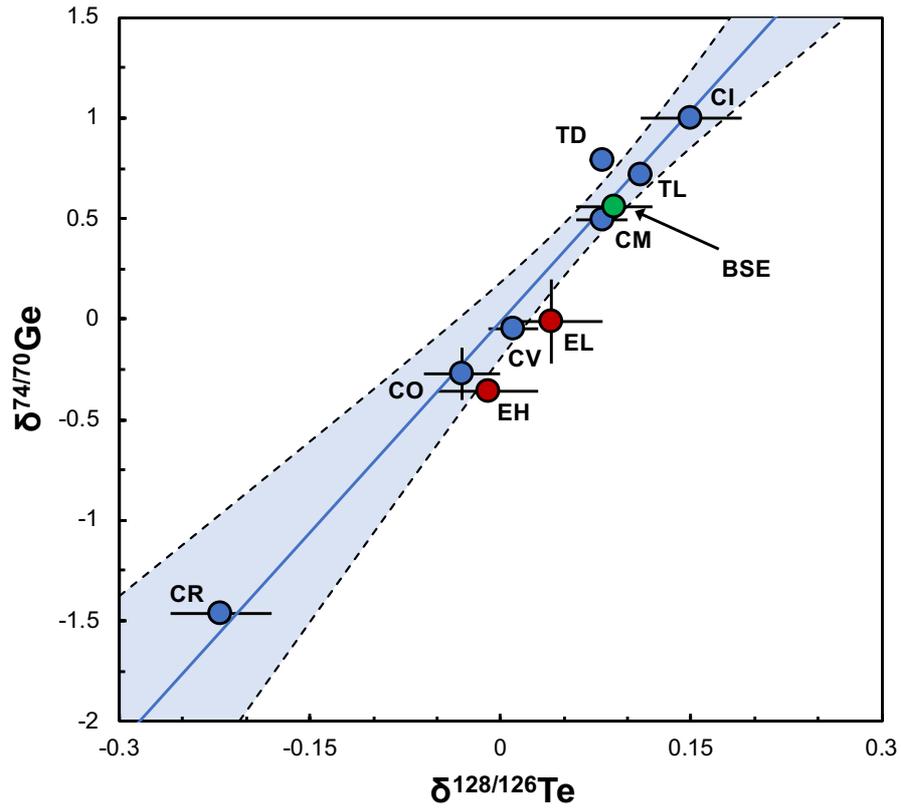

**Fig. 8:** $\delta^{128/126}$Te versus $\delta^{74/70}$Ge for carbonaceous chondrites (blue circles), enstatite chondrites (red circles), and the bulk silicate Earth (green circle). Group averages for enstatite chondrites are based solely on samples of petrological type 3 and 4. Tellurium data are from Hellmann et al. (2020) and Hellmann et al. (2021). As in Fig. 2, linear regressions were calculated using Isoplot (Ludwig, 2012). TD: Tarda. TL: Tagish Lake. The fraction of CC-derived Te in the BSE can be calculated using the lever rule [equation (1)] along the correlation line. For mixing between enstatite chondrites and CI chondrites, the fraction of CC-derived Te in the BSE is 0.57±0.40 (2σ). At this stage, the large uncertainty on this value precludes using Te isotopes as a genetic tracer for the late veneer (see section 5.3).

# Origin of moderately volatile elements in Earth inferred from mass-dependent Ge isotope variations among chondrites

Elias Wölfer, Christoph Burkhardt, Francis Nimmo, and Thorsten Kleine

**This file includes:**

Extended Methods

Discussion of the Mo isotope composition of the BSE and NC meteorites

Supplementary Figure 1

Supplementary Tables 1–2

# 1 Extended Methods

## 1.1 *Sample preparation and chemical separation of Ge*

For all samples, the (powdered) bulk rock aliquots were weighed into 15 ml Savillex PFA vials and mixed with appropriate amounts of a $^{70}$Ge–$^{73}$Ge double spike (Wölfer et al., 2025). The samples were digested on a hot-plate (120 °C, 5 days) using a (2:1) mixture of conc. HF-HNO$_3$. After digestion, the samples were evaporated and re-dissolved multiple times in concentrated HNO$_3$ at 120 °C to fume off fluorides, and to ensure complete sample dissolution and sample-spike equilibration. The samples were then dried down twice in 1 M HF and finally re-dissolved in 2 ml 1 M HF in preparation for ion exchange chromatography. Fluorides that may have formed again during this step, were centrifuged and separated.

In a first step, Ge was separated from the silicate matrices via a two-stage ion exchange chemistry broadly following previously established protocols (Luais, 2007, 2012). The samples were loaded in 2 ml 1 M HF onto Bio-Rad columns filled with 2 ml of pre-cleaned and conditioned Bio-Rad AG 1-X8 anion exchange resin (200–400 mesh), and most matrix elements were washed off by additional 13 ml 1 M HF, 2 ml H$_2$O, and 4 ml 0.2 M HNO$_3$. By contrast, Ge remained on the resin and was subsequently eluted in 20 ml 0.2 M HNO$_3$. Germanium was then separated from the remaining matrix elements using Bio-Rad columns filled with 2 ml of pre-cleaned and conditioned Bio-Rad AG 50W-X8 cation exchange resin (200–400 mesh). Sample solutions were loaded in 2 ml 0.5 M HNO$_3$ and Ge was directly eluted by additional 8 ml 0.5 M HNO$_3$, whereas the remaining matrix elements remained on the resin. The final Ge cuts were dried down, treated with concentrated HNO$_3$ to destroy organic compounds from the cation resin, and then re-dissolved in 2 ml 0.5 M HNO$_3$ for Ge isotope measurements. The Ge yields were >90% and the total procedural blank was < 1 ng and, thus, negligible for all samples. Because the Ge double spike is added prior to sample digestion, the slightly incomplete yields are inconsequential for the Ge isotope measurements.

*1.2 Isotope measurements of Ge*

The Ge isotope measurements were performed on a Thermo Scientific Neoma multicollector ICP-MS at the Max Planck Institute for Solar System Research in Göttingen, following the analytical protocol of Wölfer et al. (2025). Samples were introduced using a Cetac Aridus II desolvator and a Savillex C-flow nebulizer at an uptake rate of ~70 μl/min. Using standard sampler and X skimmer cones, a signal intensity of ~10 V on $^{70}$Ge was obtained for an optimally spiked ~100 ppb Ge solution. Each measurement consisted of on-peak zero measurements of 20 × 8 s, followed by 50 isotope ratio measurements of 8 s each. Isobaric interferences of Zn on $^{70}$Ge and of Se on $^{74}$Ge and $^{76}$Ge were corrected by monitoring $^{66}$Zn and $^{77}$Se and using the exponential mass fractionation law. Prior to each analysis, the sample introduction system was washed using 0.28 M HNO$_3$ for 5 min. Processing of the measured raw data was performed off-line following the three-dimensional data reduction scheme of Siebert et al. (2001), as described by Wölfer et al. (2025). The results are shown in the δ$^{74/70}$Ge notation as the permil deviation of the $^{74}$Ge/$^{70}$Ge ratio of a sample from the composition of the NIST SRM 3210a Ge standard:

(1) $\delta^{74/70}Ge = \left[\frac{(^{74}Ge/^{70}Ge)_{sample}}{(^{74}Ge/^{70}Ge)_{SRM3120a}} - 1\right] \times 1000$

The results are reported as the mean of replicate measurements and the corresponding errors are Student-t 95% confidence intervals (95% CI). The accuracy and reproducibility of the Ge isotope measurements were assessed by repeated analyses of the Alfa Aesar Ge solution standard as well as two terrestrial basalts standards (BHVO-2 and BCR-2), which have been processed through the full chemical separation described above and analyzed together with the chondritic samples. The measured Ge isotopic compositions of the Alfa Aesar Ge solution standard, BHVO-2, and BCR-2 are $\delta^{74/70}$Ge = –0.75 ± 0.01 (95% CI, 2s.d. = 0.07, N = 23), $\delta^{74/70}$Ge = 0.53 ± 0.03 (95% CI, 2s.d. = 0.05, N = 6), and $\delta^{74/70}$Ge = 0.58 ± 0.07 (95% CI, 2s.d. = 0.09, N = 4), respectively, and are excellent agreement with previously reported results (Escoube et al., 2012; Luais, 2012; Rouxel and Luais, 2017; Meng and Hu, 2018; Wölfer et al., 2025). Along the stable isotope measurements, precise bulk Ge concentrations of the samples were determined by isotope dilution.

## 2    Mo isotope composition of the BSE and NC meteorites

From a nucleosynthetic isotope standpoint, the late-stage delivery of CC material to the Earth is based entirely on the mixed NC-CC isotopic composition of the BSE (Budde et al., 2019). However, Bermingham et al. (2025) recently argued that the BSE's Mo is purely NC, based on the analyses of terrestrial molybdenites and a handful of iron meteorites. In particular, these authors reported new Mo isotopic data for two group IAB iron meteorites (which are NC), of which one overlaps with their revised Mo isotopic composition of the BSE. However, below we show that when all available Mo isotopic data for IAB iron meteorites are considered, the Mo isotopic composition of the BSE is distinct from that of the IAB irons.

**Fig. S1** shows Mo isotopic data for IAB irons in a plot of $\mu^{94}$Mo versus $\mu^{95}$Mo. In this diagram, all meteorites plot along two subparallel lines, which have been termed the NC- and CC-line. The contribution of NC and CC materials to the BSE's Mo can thus be determined from the position of the BSE with respect to the NC- and CC-lines. Budde et al. (2019) argued that the BSE plots between these two lines, whereas Bermingham et al. (2025) argued that the BSE plots on the NC-line. While these disparate results partly reflect differences in the BSE's

Mo isotopic composition inferred in the two studies, the central issue lies in the Mo isotopic composition of the IAB irons, which define the slope and intercept of the NC-line near the terrestrial Mo isotopic composition.

**Fig. S1** shows Mo isotopic data for IAB irons from several studies, which include data obtained by TIMS (Bermingham et al., 2025; Worsham et al., 2017) and MC-ICPMS (Marti et al., 2023; Poole et al., 2017). Most of these data plot on the NC-line defined by Spitzer et al. (2020) (solid red line) and slightly below the NC-line as defined by Bermingham et al. (2025). The Mo isotope anomalies among these IABs define a broad trend, which is consistent with variations induced by secondary neutron capture during cosmic ray exposure (CRE). These CRE effects lower both $\mu^{94}$Mo and $\mu^{95}$Mo, and so those samples affected by CRE are the samples that plot away from the NC-line. Importantly, the array of IAB data points does not pass through the BSE, regardless of which estimate for the BSE's Mo isotope composition is used. Moreover, because the CRE-unaffected IAB irons plot at the upper right end of the data array, their Mo isotopic composition is in fact quite different from that of the BSE.

The only IAB iron whose Mo isotopic composition appears to overlap with the BSE estimate from Bermingham et al. (2025) is the analyses of Campo del Cielo from Bermingham et al. (2025). But this same sample has also been analyzed in three other studies; these three studies report consistent results for Campo del Cielo and small differences among the individual data points can be accounted for by small CRE effects in some of the samples (**Fig. S1b**). However, the data point from Bermingham et al. (2025) (shown by a solid diamond in **Fig. S1**) is offset from these other data points towards lower $\mu^{94}$Mo values and as a result plots slightly to the left of the IAB array defined by the other samples. This is the reason why this single analysis of Campo del Cielo overlaps with the BSE composition of Bermingham et al. (2025). Also, in the regression to calculate the NC line, it is this sample that makes the NC line shallower and results in a more positive *y*-axis intercept. We conclude that the notion of a purely NC origin of the BSE's Mo appears to depend almost entirely on the Mo isotope data point of a single IAB iron; this data point, however, is inconsistent with all other analyses of IAB irons.

As pointed out by Budde et al. (2023), non-exponential isotope fractionation can result in spurious Mo isotope anomalies through the internal normalization of the data necessary to correction for instrumental mass fractionation. These effects are most pronounced for $^{92}$Mo, $^{94}$Mo, and $^{100}$Mo and can result in downward shifts of µ$^{92}$Mo, µ$^{94}$Mo, and µ$^{100}$Mo [see Fig. 5 of Budde et al. (2023)]. Of note, the Campo del Cielo analyses of Bermingham et al. (2025) shows negative µ$^{92}$Mo and µ$^{100}$Mo values; such compositions are unobserved among other meteorites, all of which show positive µ$^{92}$Mo values. As such, this pattern suggest that the Mo isotope anomalies reported for Campo del Cielo by Bermingham et al. (2025) may be affected by non-exponential mass fractionation as described in Budde et al. (2023). This would result in a downward shift of its µ$^{94}$Mo value and could thus explain why this single sample plots to the left of the IAB array defined by all other samples. Of note, the other IAB iron meteorite analyzed by Bermingham et al. (2025), Hope, does not show negative µ$^{92}$Mo and µ$^{100}$Mo values, and plots within the array defined by the other IAB irons. We, therefore, conclude that the Mo isotopic composition of IAB irons is distinct from the BSE and that both estimates for the BSE's Mo isotopic composition plot above the NC-line, indicating a significant contribution of CC-derived Mo.

# 3 Supplementary figures

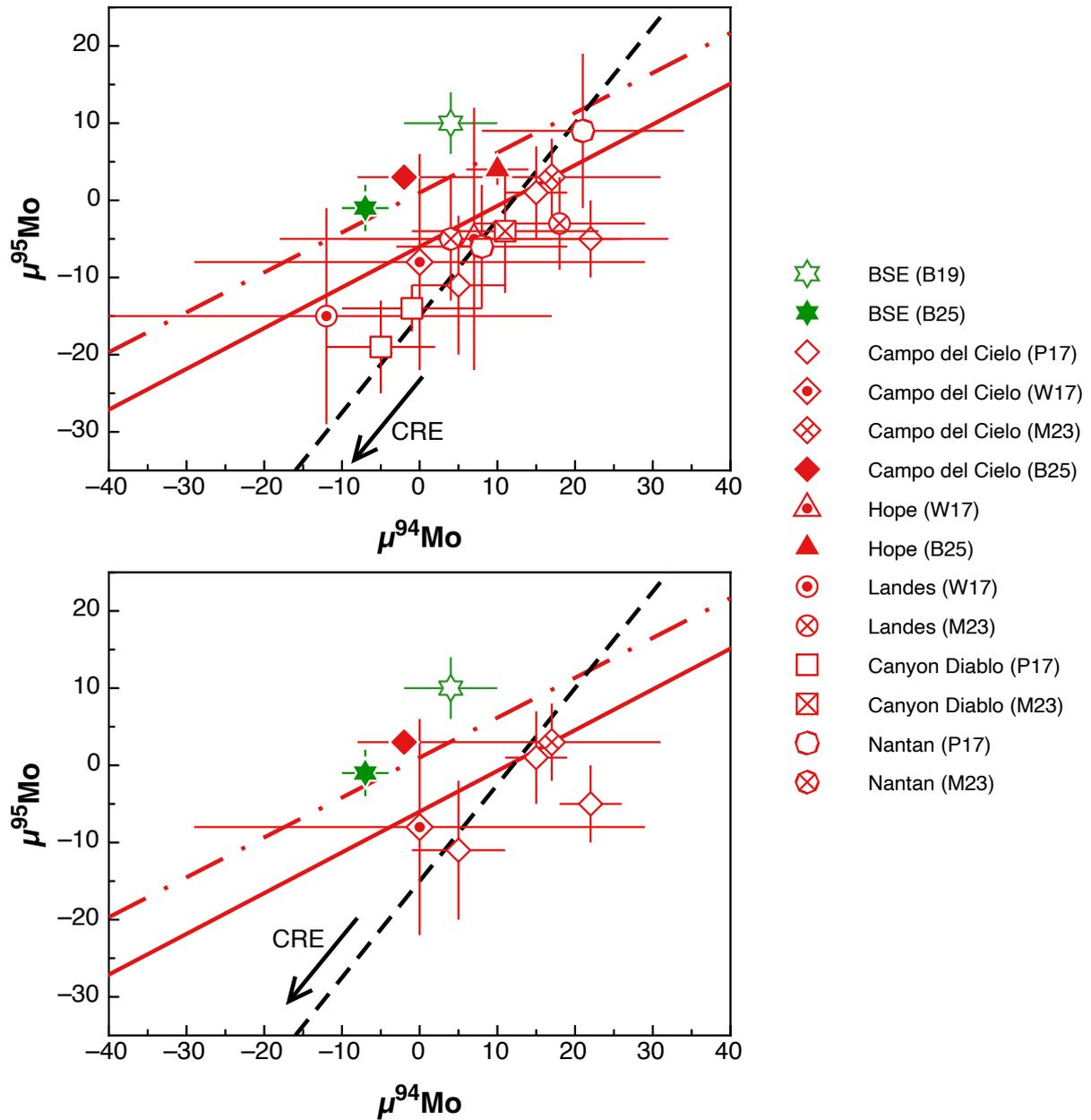

**Fig. S1**: $\mu^{94}$Mo versus $\mu^{95}$Mo for IAB iron meteorites. Two estimates for the composition of the BSE are shown for comparison. Solid red line is the NC-line from Spitzer et al. (2020); dashed-dotted line is the NC-line from Bermingham et al. (2025). Dashed black line indicates expected modifications from CRE-induced modifications of Mo isotope compositions [based on Spitzer et al. (2020)]. Data sources: P17 = Poole et al. (2017); W17 = Worsham et al. (2017); M23 = (Marti et al., 2023); B25 = Bermingham et al. (2025).

# 4 Supplementary tables

**Table S1:** Elemental and isotopic compositions of CI-like matrix and the non-matrix component.

| Element | $\delta^{XX}CI^a$ (±2σ) | $\delta^{XX}NM^b$ (±2σ) | $\Delta_{CI-NM}$/amu [‰]$^c$ (±2σ) | abundance/CI$^d$ (±2σ) | Reference |
|---|---|---|---|---|---|
| Ge | 1.00 ± 0.04 | $-2.62^{+0.87}_{-1.13}$ | $0.91^{+0.22}_{-0.28}$ | 0.25 ± 0.04 | this study |
| K  | 0.04 ± 0.08 | –0.33 ± 0.12 | 0.19 ± 0.07 | 0.48 ± 0.04 | (1) |
| Rb | 0.19 ± 0.03 | 0.04 ± 0.05 | 0.08 ± 0.03 | 0.42 ± 0.04 | (1) |
| Zn | 0.47 ± 0.03 | 0.05 ± 0.09 | 0.21 ± 0.05 | 0.19 ± 0.03 | (1) |
| Te | 0.15 ± 0.01 | –0.38 ± 0.09 | 0.27 ± 0.05 | 0.15 ± 0.02 | (1) |
| Cd | 0.38 ± 0.17 | –0.19 ± 0.27 | 0.14 ± 0.08 | 0.11 ± 0.06 | (2) |

$^a$ Mass-dependent isotopic composition of the CI-like matrix (i.e., CI chondrites).

$^b$ Mass-dependent isotopic composition of the inferred non-matrix component.

$^c$ Magnitude of isotopic fractionation per amu between CI-like matrix and the non-matrix component.

$^d$ CI-normalized elemental abundance in the non-matrix component.

References: (1) Nie et al. (2021), (2) Morton et al. (2024).

**Table S2:** Average Ge isotopic compositions of different planetary materials.

| Meteorite group | $\delta^{74}Ge$ (±2σ) | Reference |
|---|---|---|
| CI | 1.00 ± 0.04 | this study |
| TL | 0.71 ± 0.04 | this study |
| TD | 0.79 ± 0.04 | this study |
| CM | 0.49 ± 0.02 | this study |
| CO | –0.27 ± 0.13 | this study |
| CV | –0.05 ± 0.04 | this study |
| CR | –1.47 ± 0.04 | this study |
| IIC | 1.30 ± 0.09 | (1–2) |
| EH | –0.36 ± 0.05 | this study |
| EL | –0.01 ± 0.21 | this study |
| H | –0.51 ± 0.09 | (3) |
| L | –0.31 ± 0.06 | (3) |
| LL | –0.26 ± 0.08 | (3) |
| IIAB/IIG | 1.17 ± 0.18 | (1–2) |
| IIIAB | 1.55 ± 0.39 | (1–2) |
| IAB | 0.80 ± 0.09 | (1–2), (4) |
| BSE | 0.60 ± 0.02 | (1–2), (4–7) |

The $\delta^{74/70}Ge$ data of the groups represent the weighted mean of the individual samples of each group, calculated using IsoplotR. The individual data from each group are from the following references: (1) (Luais, 2007), (2) (Luais, 2012), (3) (Florin et al., 2020), (4) (Escoube et al., 2012), (5) (Rouxel et al., 2006), (6) (Rouxel and Luais, 2017), (7) (Meng and Hu, 2018).

# 5 Supplementary references